\documentclass[]{article}

\usepackage{paralist}

\usepackage{calrsfs}

\usepackage{oldgerm}

\usepackage{subeqn}

\usepackage[dvips]{graphicx}

\usepackage{amssymb,amsbsy}
\usepackage{epic,eepic}
\usepackage{texdraw}
\usepackage{dsfont}
\usepackage{latexsym}
\usepackage{ntheorem}

\usepackage{accents}
\DeclareMathAccent{\wtilde}{\mathord}{largesymbols}{"65}
\DeclareMathAccent{\what}{\mathord}{largesymbols}{"62}

\makeatletter
\def\wb{\accentset{{\cc@style\underline{\mskip10mu}}}}
\makeatother

\newcommand{\be}{\begin{equation}}
\newcommand{\ee}{\end{equation}}
\newcommand{\bdm}{\begin{displaymath}}
\newcommand{\edm}{\end{displaymath}}

\theoremstyle{break}

\newtheorem{thm}{Proposition}[section]

\theorembodyfont{\upshape}

\newtheorem{rem}{Remark}[section]
\newtheorem*{pr}{Proof}

\makeatother

\begin{document}
\title{Symmetries and integrability of discrete equations\\ defined on a black--white lattice}

\author{P D Xenitidis and V G Papageorgiou\\
Department of Mathematics, University of Patras, 265 00 Patras, Greece \\
{\footnotesize{e--mail: xeniti@math.upatras.gr, vassilis@math.upatras.gr}}}

\maketitle

\begin{abstract}
We study the deformations of the H equations, presented recently by Adler, Bobenko and Suris, which are naturally defined on a black-white lattice. For each one of these equations, two different three-leg forms are constructed, leading to two different discrete Toda type equations. Their multidimensional consistency leads to B{\"a}cklund transformations relating different members of this class, as well as to Lax pairs. Their symmetry analysis is presented yielding infinite hierarchies of generalized symmetries.
\end{abstract}


\section{Introduction}

Recently, Adler, Bobenko and Suris using a general setup studied affine linear quad-equations, which are multidimensionally consistent, \cite{ABS1}. As an outcome, they extended the lists of the equations given in \cite{ABS} by presenting {\it{deformations of the}} H {\it{equations}}. These equations differ from the original H equations in \cite{ABS}, since they possess the symmetries of the rhombus instead of the square, and are naturally defined on a black--white lattice.

In this paper, we study some properties of the above equations. Specifically, we derive two different three-leg forms for each one of the equations under consideration. The existence of two such forms is justified by the dependence of the deformed equations on black and white points. As a consequence, we construct two different discrete Toda equations from each equation in this class. Using the multidimensional consistency of the deformed equations, we construct a B{\"a}cklund transformation relating two different members of this class. This transformation is also employed in the derivation of Lax pairs for each one of these equations.

Another interesting aspect of the deformed H equations is the structure of their symmetries. Their symmetry analysis is similar to the one presented in \cite{TTX} and \cite{X1}, and implies that, each one of the deformed H equations admits a pair of generalized symmetries and a pair of extended symmetries, as well. Using both of these pairs of symmetries, we derive infinite hierarchies of generalized symmetries.

The paper is organized as follows. In section 2, we introduce the notation as well as preliminaries on the symmetries of lattice equations. The next section is devoted to the presentation of the deformations of the H equations and their relation to Toda type systems. Their integrability aspects, i.e. B{\"a}cklund transformations and Lax pairs, are given in section 4, while their symmetry analysis is contained in the following section. Various perspectives are contained in section 6.

\section{Notation and preliminaries on symmetries of discrete equations}

In this section, we provide notation and termilology of symmetries of difference equations to be used in what follows. A detailed presentation of symmetries of difference equations can be found in \cite{tp:Levi1a}.

A partial difference equation is a functional relation among the values of a function $u : {\mathds{Z}} \times {\mathds{Z}} \rightarrow {\mathds{C}}$ at different points of the lattice, which may involve the independent variables $n$, $m$ and the lattice spacings $\alpha$, $\beta$, as well, i.e. a relation of the form
\be {\cal E}(u_{n,m},u_{n+1,m},u_{n,m+1},\ldots;n,m;\alpha,\beta)\,=\,0\,. \label{gendisceq} \ee
In this relation, $u_{n,m}$ is the value of the function $u$ at the lattice point $(n,m)$, i.e. $u_{n,m}\,=\,u(n,m)$, and this is the notation that we will adopt for the values of the function $u$ from now on.

The analysis of these equations is facilitated by the two translation (or shift) operators acting on functions on ${\mathds{Z}}^2$, which are defined by 
$$\left( \mathcal{S}_n^{(k)} u \right)_{n,m} = u_{n+k,m}\,,\quad \left( \mathcal{S}_m^{(k)} u\right)_{n,m} = u_{n,m+k}\,,\quad {\mbox{where}}\,\, k \in \mathds{Z} \,.$$

Let $\mbox{G}$ be a connected one-parameter group of transformations acting on the domain of the dependent variable $u_{n,m}$ of the lattice equation (\ref{gendisceq}), i.e.
$${\mbox{G}}\,:\,u_{n,m}\,\longrightarrow\,{\tilde u}_{n,m}\,=\,\Phi(n,m,u_{n,m};\varepsilon)\,,\quad \varepsilon \,\,\in\,\,{\mathds{R}}\,. $$
The prolongation of the group action of $\rm G$ on the shifted values of $u$ is defined by
\begin{equation}
{\rm G}^{(k)}\,:\,(u_{n+i,m+j}) \,\longrightarrow\,\left({\tilde u}_{n+i,m+j}\,=\,\Phi(n+i,m+j,u_{n+i,m+j};\varepsilon)\right)\,.
\label{eq:gract} \end{equation}

The transformation group ${\rm G}$ is a {\it{local Lie point symmetry}} of the lattice equation (\ref{gendisceq}) if it transforms any solution of (\ref{gendisceq}) to another solution of the same equation. The {\it{infinitesimal criterion}} for ${\rm G}$ to be a symmetry  of equation (\ref{gendisceq}) is
\begin{equation}
 {\mathbf{x}}^{(k)} \left({\mathcal{E}}\left(u_{n,m},u_{n+1,m},u_{n,m+1},\ldots;n,m;\alpha,\beta\right) \right) \,=\,0\,,
\label{eq:infcr} \end{equation}
which should hold for every solution of equation (\ref{gendisceq}). In the above relation, the vector field ${\mathbf{x}}\,=\,R(n,m,u_{n,m})\,\partial_{u_{n,m}}$ with its {\it{characteristic}} defined by 
$$R(n,m,u_{n,m})\,=\, \left . \frac{{\rm{d}} \phantom{\varepsilon}}{{\rm{d}} \varepsilon} \Phi(n,m,u_{n,m};\varepsilon) \right| _{\varepsilon = 0}\,,$$
is the {\emph{infinitesimal generator}} of the group action of $\rm G$, and
$${\mathbf{x}}^{(k)}\,=\,\sum_{i=0}^{k}\sum_{j=0}^{k-i} \left({\mathcal{S}}_n^{(i)}\circ {\mathcal{S}}_m^{(j)} R\right)(n,m,u_{n,m}) \,\partial_{u_{n+i,m+j}}$$
is its $k$-th order forward prolongation.

By extending the geometric transformations to more general ones, which depend, not only on $n$, $m$ and $u_{n,m}$, but also on the shifted values of $u$, we arrive naturally at the notion of {\emph{generalized symmetry}}. In this case, the characteristic of the infinitesimal generator is a function of the form $R(n,m,u_{n,m},u_{n+1,m},u_{n-1,m},u_{n,m+1},\ldots)$.

A further generalization of symmetries follows by considering transformations acting on $u$ and the lattice parameters, as well. We will refer to them as {\emph{extended symmetries}}. Such symmetries can be used effectively for the construction of similarity solutions, \cite{TX}, and derivation of higher order generalized symmetries, \cite{X1}.

\section{Integrable discrete equations on a black-white lattice}

In this section we first present some general characteristics of a class $\cal{G}$ of lattice equations possessing the symmetries of the rhombus. Next, we study the equations presented in \cite{ABS1}, which possess the rhombic symmetry but are defined on a black--white lattice. Finally, we derive some of the properties of the latter equations, as well as their relations to discrete Toda type systems.

\begin{figure}[ht]
\begin{minipage}{16	pc}
\centertexdraw{ \setunitscale 0.5
\linewd 0.02 \arrowheadtype t:F 
\htext(0 0.5) {\phantom{T}}
\move (-1 -2) \lvec (1 -2) 
\move(-1 -2) \lvec (-1 0) \move(1 -2) \lvec (1 0) \move(-1 0) \lvec(1 0)
\move (1 -2) \fcir f:0.0 r:0.1 \move (-1 -2) \fcir f:0.0 r:0.1
 \move (-1 0) \fcir f:0.0 r:0.1 \move (1 0) \fcir f:0.0 r:0.1  
\htext (-1.1 -2.5) {$u_{n,m}$} \htext (.9 -2.5) {$u_{n+1,m}$} \htext (0 -2.25) {$\alpha$}
\htext (-1.1 .15) {$u_{n,m+1}$} \htext (.9 .15) {$u_{n+1,m+1}$} \htext (0 .1) {$\alpha$}
\htext (-1.25 -1) {$\beta$} \htext (1.1 -1) {$\beta$}}
\caption{{\em{The quadrilateral}}} \label{fig:quad}
\end{minipage}%
\begin{minipage}{16pc}
\centertexdraw{ \setunitscale 0.5
\linewd 0.02 \arrowheadtype t:F 
\htext(0 0.5) {\phantom{T}}
\move (-1 -2) \lvec (1 -2) 
\move(-1 -2) \lvec (-1 0) \move(1 -2) \lvec (1 0) \move(-1 0) \lvec(1 0)
\move (1 -2) \fcir f:0.0 r:0.1 \move (-1 -2) \fcir f:0.0 r:0.1
 \move (-1 0) \fcir f:0.0 r:0.1 \move (1 0) \fcir f:0.0 r:0.1  
 \move (-1 -2) \lpatt (.1 .1) \lvec (1 0) \move (1 -2) \lvec (-1 0) \lpatt ()
\htext (-1.1 -2.5) {$u_{n,m}$} \htext (.9 -2.5) {$u_{n+1,m}$} \htext (0 -2.35) {$h_{34}$}
\htext (-1.1 .15) {$u_{n,m+1}$} \htext (.9 .15) {$u_{n+1,m+1}$} \htext (0 .1) {$h_{12}$}
\htext (-1.45 -1) {$h_{24}$} \htext (1.1 -1) {$h_{13}$}
\htext (-.5 -1.75) {$h_{23}$} \htext (.15 -1.75 ) {$h_{14}$}}
\caption{{\em{The polynomials}}} \label{polyhg}
\end{minipage}
\end{figure}

Let us first introduce the class $\cal{G}$. It contains all the autonomous discrete equations, which involve the values of a function $u$ at the vertices of an elementary quadrilateral, as shown in Figure \ref{fig:quad}, i.e. they have the form
\begin{equation}
Q(u_{n,m},u_{n+1,m},u_{n,m+1},u_{n+1,m+1};\alpha,\beta) \,=\, 0. \label{eq:genform}
\end{equation}
Furthermore, the function $Q$ satisfies the following requirements. It is affine linear, depends explicitly on the four indicated values of $u$, and possesses the symmetries of the rhombus:
\begin{eqnarray*}
 Q(u_{n,m},u_{n+1,m},u_{n,m+1},u_{n+1,m+1};\alpha,\beta) &=&  \tau \,Q(u_{n,m},u_{n,m+1},u_{n+1,m},u_{n+1,m+1};\beta,\alpha) \\
 &=&  \tau^\prime \, Q(u_{n+1,m+1},u_{n+1,m},u_{n,m+1},u_{n,m};\beta,\alpha),
\end{eqnarray*}
where $\tau = \pm 1$ and $\tau^\prime = \pm 1$.

The affine linearity of $Q$ implies that one can define six different polynomials in terms of the function $Q$, \cite{ABS,ABS1}, as indicated in Figure \ref{polyhg}. They are defined by the relations
$$h_{ij}\,=\,h_{j\,i}\,:=\,Q\,Q_{,i j}\,-\,Q_{,i}\,Q_{,j}\,,\quad i\,\ne\,j\,,\quad i,\,j\,=\,1,\ldots,\,4,$$
where $Q_{,i}$ denotes the derivative of $Q$ with respect to its $i$-th argument and $Q_{,i j}$ the second order derivative $Q$ with respect to its $i$-th and $j$-th argument.

Taking into account the rhombic symmetry of $Q$, one arrives at the following conclusions.
\begin{enumerate}
\item All the polynomials assigned to the edges are given in terms of a quadratic polynomial $h$. Specifically, the $h_{i j}$'s assigned to the horizontal edges read as
$$
h_{34} \,=\, h(u_{n,m},u_{n+1,m};\alpha,\beta)\,, \quad h_{12} \,=\, h(u_{n+1,m+1},u_{n,m+1};\alpha,\beta)\,,
$$
while the ones assigned to the vertical edges have the form
$$
h_{24} \,=\, h(u_{n,m},u_{n,m+1};\beta,\alpha)\,,\quad h_{13} \,=\, h(u_{n+1,m+1},u_{n+1,m};\beta,\alpha)\,.
$$
\item The diagonal polynomials
$$ h_{14}\,=\,G_1(u_{n+1,m},u_{n,m+1};\alpha,\beta)\,,\quad h_{23}\,=\, G_2(u_{n,m},u_{n+1,m+1};\alpha,\beta)\,, $$
are quadratic and symmetric in the first two of their arguments and symmetric in the parameters.
\end{enumerate}

\begin{figure}[th]
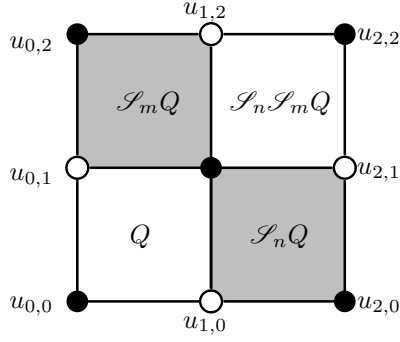

\centering
\centertexdraw{ \setunitscale .7 \linewd 0.02 \arrowheadtype t:F
\move (0 -1) \lvec (0 0) \lvec(1 0) \lvec(1 -1) \lvec(0 -1) \ifill f:0.75 
\move(-1 0) \lvec(-1 1) \lvec(0 1) \lvec(0 0) \lvec(-1 0) \ifill f:0.75 
\move(-.92 -1) \lvec(-.08 -1)
\move(-1 -.92) \lvec(-1 -0.08)
\move(-1 .1) \lvec(-1 1)
\move(-.9 0) \lvec(.9 0)
\move(0 0) \lvec(0 -.9)
\move(.1 -1) \lvec(1 -1) \lvec(1 -.1)
\move(1 .1) \lvec(1 1) \lvec(.1 1)
\move(-.1 1) \lvec (-1 1)
\move(0 .9) \lvec (0 -.9) 
\move(-1 -1) \fcir f:0.0 r:0.08
\move(0 -1) \lcir  r:0.08 \move(0 -1) \fcir f:1  r:0.07
\move(1 -1) \fcir f:0.0 r:0.08
\move(-1 0) \lcir r:0.08 \move(-1 0) \fcir f:1 r:0.07
\move(0 0) \fcir f:0.0 r:0.08
\move(1 0) \lcir r:0.08 \move(1 0) \fcir f:1 r:0.07
\move(-1 1) \fcir f:0.0 r:0.08
\move(0 1) \lcir r:0.08 \move(0 1) \fcir f:1 r:0.07
\move(1 1) \fcir f:0.0 r:0.08
\htext (-0.6 -.6) {$Q$} \htext (0.3 -.6) {${\cal{S}}_n Q$} 
\htext (-0.7 .4) {${\cal{S}}_m Q$} \htext (0.15 .4) {${\cal{S}}_n {\cal{S}}_m Q$} 
\htext (1.1 -0.1) {${u_{2,1}}$} \htext (1.1 0.9) {${u_{2,2}}$}
\htext (-0.2 1.1) {$u_{1,2}$} 
\htext (-1.5 -0.15) {$u_{0,1}$} \htext (-1.5 0.85) {$u_{0,2}$}
\htext (-1.5 -1.1) {$u_{0,0}$} \htext (1.1 -1.1) {${u_{2,0}}$} \htext (-0.2 -1.25) {${u_{1,0}}$}}
\caption{Black-white lattice and the corresponding equations} \label{rhombeq}
\end{figure}

Among the equations, which are affine linear and possess the symmetries of the rhombus, are {\it{the deformations of the {\rm H} equations}} presented recently by Adler, Bobenko and Suris in \cite{ABS1}. These equations are defined on a {\it{black--white lattice}} (or {\it{chessboard lattice}}), as indicated in Figure \ref{rhombeq}, implying that, differently colored quadrilaterals carry different equations.

Specifically, an equation of this class has the form ${\bf{Q}}[u]=0$, where either
\begin{subequations} \label{bweq}
\be
 {\bf{Q}}[u] \,:=\,\left\{\begin{array}{ll} Q(u_{n,m},u_{n+1,m},u_{n,m+1},u_{n+1,m+1};\alpha,\beta), & |n|+|m| = 2 k \\  & \\ Q(u_{n+1,m},u_{n,m},u_{n+1,m+1},u_{n,m+1};\alpha,\beta), & |n|+|m| = 2 k+1 \end{array} \right.\,, \label{bweq1}
\ee
or
\be
 {\bf{Q}}[u] \,:=\,\left\{\begin{array}{ll} Q(u_{n,m},u_{n+1,m},u_{n,m+1},u_{n+1,m+1};\alpha,\beta) , & |n|+|m| = 2 k+1 \\  & \\ Q(u_{n+1,m},u_{n,m},u_{n+1,m+1},u_{n,m+1};\alpha,\beta), & |n|+|m| = 2 k \end{array} \right.\,. \label{bweq2}
\ee
\end{subequations}
In the above relations, $k$ is a non-negative integer and the function $Q$ has one of the following forms\footnote{Compared to the equations presented in \cite{ABS1}, we have changed the sign of the parameter $\epsilon$ .}.
\begin{subequations} \label{H13ae}
\begin{enumerate}[i)]
\item Deformation of equation ${\rm H1}$
\be  Q(u,x,y,z;\alpha,\beta) = (u-z) (x-y)\, -\,(\alpha \,- \, \beta) \,\left(1\,-\,\epsilon x y \right) \label{H1ae1} \ee

\item Deformation of equation ${\rm H2}$
\begin{eqnarray}
 Q(u,x,y,z;\alpha,\beta) &&=  (u-z) (x-y) + (\beta-\alpha) (u+x+y+z) - \alpha^2 + \beta^2  \nonumber \\
 &&\quad  -\, \epsilon \,(\beta-\alpha) \,(2 x + \alpha+\beta) \,(2 y +\alpha+\beta) \,-\, \epsilon\, (\beta-\alpha)^3  \label{H2ae1}
\end{eqnarray}

\item Deformation of equation ${\rm H3}$
\be
 Q(u,x,y,z;\alpha,\beta) = \alpha (u x + y z) - \beta (u y + x z) + (\alpha^2-\beta^2) \left(\delta \,-\, \frac{\epsilon x y}{\alpha \beta} \right) \label{H3ae1}
\ee
\end{enumerate}
\end{subequations}

Combining relations (\ref{bweq}) with (\ref{H13ae}), the deformations of the H equations can be written in the following compact, but non-autonomous form, by using the functions
\begin{equation}
{\cal{X}}_{n,m}\,=\,\frac{1 + \sigma (-1)^{n+m}}{2}\,,\quad {\cal{Y}}_{n,m}\,=\,\frac{1 - \sigma (-1)^{n+m}}{2}\,, \quad \sigma \,=\,\pm 1\,.
\end{equation}
\begin{enumerate}[\rm i.]
\item Equations ${\rm H1}^\sigma_\epsilon$
\begin{eqnarray}
 &&(u_{n,m}-u_{n+1,m+1})\, (u_{n+1,m}-u_{n,m+1})\, -\,(\alpha \,- \, \beta) + \nonumber \\
 && \epsilon (\alpha - \beta)\big\{{\cal{X}}_{n,m} \,u_{n+1,m} u_{n,m+1} + {\cal{Y}}_{n,m} \,u_{n,m} u_{n+1,m+1}\big\}= 0 \label{H1ae}
\end{eqnarray}
\item Equations ${\rm H2}^\sigma_\epsilon$
\begin{eqnarray}
 && (u_{n,m}-u_{n+1,m+1})(u_{n+1,m}-u_{n,m+1}) +\nonumber \\
 && (\beta-\alpha) (u_{n,m}+u_{n+1,m}+u_{n,m+1}+u_{n+1,m+1}) - \alpha^2 + \beta^2 -  \epsilon\, (\beta-\alpha)^3  - \nonumber \\
  &&\epsilon \,(\beta-\alpha) \,\left\{ 2 {\cal{Y}}_{n,m} u_{n,m} + 2 {\cal{X}}_{n,m} u_{n+1,m}+\alpha+\beta\right\} \times \nonumber \\
 && \qquad \qquad \left\{  2 {\cal{Y}}_{n,m} u_{n+1,m+1} + 2 {\cal{X}}_{n,m} u_{n,m+1} +\alpha+\beta\right \}\,=\,0 \label{H2ae}
\end{eqnarray}
\item Equations ${\rm H3}^\sigma_\epsilon$
\begin{eqnarray}
 \alpha (u_{n,m} u_{n+1,m}+u_{n,m+1} u_{n+1,m+1}) - \beta (u_{n,m} u_{n,m+1}+u_{n+1,m} u_{n+1,m+1}) \nonumber \\
 + (\alpha^2-\beta^2) \delta  -\, \frac{\epsilon (\alpha^2-\beta^2)}{\alpha \beta}\big\{{\cal{X}}_{n,m} \,u_{n+1,m} u_{n,m+1} + {\cal{Y}}_{n,m} \,u_{n,m} u_{n+1,m+1}\big\} = 0 \label{H3ae}
\end{eqnarray}
\end{enumerate}
The equations of the form (\ref{bweq1}) correspond to the choice $\sigma=+1$, while the ones of the form (\ref{bweq2}) correspond to $\sigma=-1$.

\begin{rem}
The analysis of the deformations of the H equations can be easily performed using their forms (\ref{bweq}) and treating them as members of the class $\cal{G}$. However, we find it more convenient to use their non-autonomous expressions (\ref{H1ae})-(\ref{H3ae}) in order to present the corresponding results in a more compact form. Thus, in the following, we will denote the left hand side of equations (\ref{H1ae})-(\ref{H3ae}) by \\ ${\cal{E}}(u_{n,m},u_{n+1,m},u_{n,m+1},u_{n+1,m+1};n,m;\alpha,\beta;\sigma)$. \hfill $\Box$
\end{rem}

\begin{rem}
Equations ${\rm H}^{\sigma}_\epsilon$ corresponding to the two different choices of $\sigma$ are counterparts, since they correspond to alternative coloring of vertices, cf. Figure \ref{correspond}, i.e.
\begin{eqnarray*}
 && {\cal{E}}(u_{n,m},u_{n+1,m},u_{n,m+1},u_{n+1,m+1};n,m;\alpha,\beta;-\sigma) \,=\, \\
 && \qquad\qquad\qquad {\cal{E}}(u_{n+1,m},u_{n,m},u_{n+1,m+1},u_{n,m+1};n,m;\alpha,\beta;\sigma)\,.
\end{eqnarray*}
In what follows, we will consistently use and refer to the coloring convention employed in Figure \ref{correspond}. Specifically, if $\sigma =1$ then $(0,0)$ will be a black point, and if $\sigma = -1$ then $(0,0)$ will be a white point.

\begin{figure}[h]
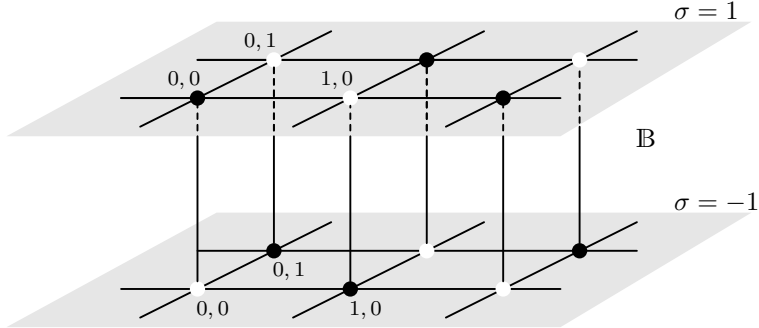

\centertexdraw{ \setunitscale 1 \linewd 0.01 \arrowheadtype t:F \arrowheadsize l:0.2 w:0.16
\move(-.3 -.05) \lvec(2.6 -.05) \lvec(3.6 .55) \lvec(.8 .55) \lvec(-.3 -.05) \ifill f:.9
\move(-.3 .95) \lvec(2.6 .95) \lvec(3.6 1.55) \lvec(.8 1.55) \lvec(-.3 .95) \ifill f:.9

\move(2 0) \lvec(3 .5) 
\move(2 1)  \lvec(3 1.5) 

\move(.4 0) \lvec(1.4 .5) 
\move(1.2 0) \lvec(2.2 .5) 
\move(.4 1)  \lvec(1.4 1.5) 
\move(1.2 1) \lvec(2.2 1.5) 
\move(.3 .15) \lvec(2.3 .15) \move(.7 .35) \lvec(2.7 .35) 
\move(.3 1.15) \lvec(2.3 1.15) \move(.7 1.35) \lvec(2.7 1.35) 
\move(.3 1.15) \lvec(2.3 1.15) \move(.7 1.35) \lvec(2.7 1.35)
\move(2.3 .15) \lvec(2.6 .15)  \move(2.7 .35) \lvec(3 .35) 
\move(2.3 1.15) \lvec(2.6 1.15)  \move(2.7 1.35) \lvec(3 1.35) 

\lpatt( )
\move(.7 .15) \fcir f:1 r:0.04 
\move(1.5 .15) \fcir f:.0 r:0.04 
\move(2.3 .15) \fcir f:1 r:0.04  \move(2.3 1.15) \fcir f:.0 r:0.04 
\move(1.1 .35) \fcir f:.0 r:0.04 
\move(1.9 .35) \fcir f:1 r:0.04 
\move(2.7 .35) \fcir f:.0 r:0.04 \move(2.7 1.35) \fcir f:1 r:0.04
\move(.7 1.15) \fcir f:.0 r:0.04 
\move(1.5 1.15) \fcir f:1 r:0.04 
\move(1.1 1.35) \fcir f:1 r:0.04 
\move(1.9 1.35) \fcir f:.0 r:0.04 

\move(.7 .19) \lvec(.7 .95) \move(1.5 .19) \lvec(1.5 .95)
\move(1.1 .39) \lvec(1.1 .95) \move(1.9 .39) \lvec(1.9 .95)
\move(2.3 .19) \lvec(2.3 .95) \move(2.7 .39) \lvec(2.7 1)

\lpatt(.03 .03 )
\move(.7 .98) \lvec(.7 1.15) \move(1.5 .98) \lvec(1.5 1.11) 
\move(1.1 .98) \lvec(1.1 1.31) \move(1.9 .98) \lvec(1.9 1.31)
\move(1.1 .98) \lvec(1.1 1.31) \move(1.9 .98) \lvec(1.9 1.31)
\move(2.3 .98) \lvec(2.3 1.15) \move(2.7 1.02) \lvec(2.7 1.31)

\move(.7 .0) \htext{{\footnotesize{$0,0$}}}
\move(1.1 .2) \htext{{\footnotesize{$0,1$}}}
\move(1.5 .0) \htext{{\footnotesize{$1,0$}}}
\move(.55 1.2) \htext{{\footnotesize{$0,0$}}}
\move(.95 1.4) \htext{{\footnotesize{$0,1$}}}
\move(1.35 1.2) \htext{{\footnotesize{$1,0$}}}

\move(3.2 .56) \htext{$\sigma = -1$} \move(3.2 1.57) \htext{$\sigma = 1$} \move(3. .9) \htext{$\mathds{B}$}}
\caption{The correspondence of the ${\rm H}_\epsilon^\sigma$ equations and their multidimensional consistency} \label{correspond}
\end{figure}

Additionally, equations ${\rm H}^{\sigma}_\epsilon$ are related to each other through a B{\"a}cklund transformation $\mathds{B}$, which is given in the next section. Moreover, independently of the choice of $\sigma$, equations (\ref{H1ae})-(\ref{H3ae}) reduce to the corresponding H equations of \cite{ABS}, when the additional parameter $\epsilon$ is set equal to 0. Thus, in the following, the former equations will also be referred to as the {\it{deformed equations}}. \hfill $\Box$
\end{rem}

Equations ${\rm H}_\epsilon^\sigma$ are multidimensionally consistent and possess the tetrahedron property, \cite{ABS1}. These two properties imply that, the polynomial $h$ is factorized as
\be h(x,y;n,m;\alpha,\beta;\sigma)\,=\,k(\alpha,\beta)\,f(x,y;n,m;\alpha;\sigma)\,, \label{hcomp} \ee
where $f$ is a biquadratic polynomial of $x$ and $y$, and the function $k$ is antisymmetric, i.e. $k(\beta,\alpha)\,=\,-\,k(\alpha,\beta)$. In the following, we will omit the dependence of $f$ on $n$, $m$ and $\sigma$, in order to simplify our notation. The polynomials $f$, $G_1$, $G_2$ and $k$ related to the deformed equations are given in the Appendix, cf. also \cite{ABS1}. 

The tetrahedron property implies the existence of a {\it{three--leg form}} of all of the deformed equations, for instance a relation of the form\footnote{We present its multiplicative formulation, because it turns out that, in almost all cases it is more convenient to write the three-leg form in this fashion.}
$$\frac{F(u_{0,0},u_{1,0},\alpha)}{F(u_{0,0},u_{0,1},\beta)} \,=\,W(u_{0,0},u_{1,1})\quad {\mbox{or}} \quad \frac{F(u_{1,0},u_{0,0},\alpha)}{F(u_{1,0},u_{1,1},\beta)} \,=\,W(u_{1,0},u_{0,1})\,,$$
where $F$ and $W$ correspond to the edges (``short'' legs) and the diagonals (``long'' legs) of an elementary quadrilateral, respectively.

The three-leg form depends essentially on the ``color'' of the vertices of the ``long'' leg, cf. Figure \ref{correspond}. Thus, we come up with two different three--leg forms for each one of the deformed equations, which are given in the following list.
\begin{enumerate}
\item Equations ${\rm H1}_\epsilon^\sigma$\\
{\it{White diagonal}}: additive three-leg form $F(x_1,x_2,\alpha)-F(x_1,x_3,\beta)=W(x_1,x_4)$
$$F(x_1,x_2,\alpha)\,=\,x_1 + x_2\,,\quad W(x_1,x_4)\,=\,(\alpha-\beta)\,\frac{1-\epsilon\, x_1 x_4}{x_1-x_4} $$
{\it{Black diagonal}}
$$F(y_1,y_2,\alpha)\,=\,\frac{1 - \sqrt{\epsilon} y_1}{1 + \sqrt{\epsilon} y_1}\,\frac{1 - \sqrt{\epsilon} y_2}{1 + \sqrt{\epsilon} y_2}\,,\quad W(y_1,y_4)\,=\,\frac{y_4-y_1 + \sqrt{\epsilon} (\alpha-\beta)}{y_4-y_1 - \sqrt{\epsilon} (\alpha-\beta)}$$

\item Equations ${\rm H2}_\epsilon^\sigma$\\
{\it{White diagonal}}
$$F(x_1,x_2,\alpha)\,=\,x_1 + x_2 + \alpha - 2 \epsilon (x_1+\alpha)^2\,,\quad W(x_1,x_4)\,=\,\frac{x_1 -x_4+\alpha-\beta}{x_1 -x_4-\alpha+\beta}$$
{\it{Black diagonal}}
$$F(y_1,y_2,\alpha)\,=\,\frac{1-4 \,\epsilon \,(y_2 + \alpha) - \sqrt{1 + 8 \epsilon y_1}}{1-4\, \epsilon\, (y_2+\alpha)  + \sqrt{1 + 8 \epsilon y_1}}\,,$$
$$W(y_1,y_4)\,=\, \frac{y_1-y_4 + (\alpha-\beta) (\,2 \epsilon (\alpha-\beta) + \sqrt{1 + 8 \epsilon y_1}\,)}{y_1-y_4 + (\alpha-\beta) (\,2 \epsilon (\alpha-\beta) - \sqrt{1 + 8 \epsilon y_1}\,)}$$

\item Equations ${\rm H3}_\epsilon^\sigma$\\
{\it{White diagonal}}
$$F(x_1,x_2,\alpha)\,=\,\alpha\,x_1 x_2 + \delta\, \alpha^2 - \epsilon\, x_1^2\,,\quad W(x_1,x_4)\,=\,\frac{\alpha}{\beta}\,\frac{\alpha x_4 - \beta x_1}{\beta x_4 - \alpha x_1}$$
{\it{Black diagonal}}
$$F(y_1,y_2,\alpha)\,=\,\frac{\alpha (\,y_1 + \sqrt{y_1^2 + 4 \epsilon \delta}\,)-2 \epsilon y_2}{\alpha (\,y_1 + \sqrt{y_1^2 + 4 \epsilon \delta}\,) + 2 \epsilon y_2}\,,$$
$$W(y_1,y_4)\,=\, \frac{(\alpha^2+\beta^2) y_1 - 2 \alpha \beta y_4 + (\alpha^2-\beta^2) \sqrt{y_1^2 + 4 \epsilon \delta}}{(\alpha^2+\beta^2) y_1 - 2 \alpha \beta y_4 - (\alpha^2-\beta^2) \sqrt{y_1^2 + 4 \epsilon \delta}}$$
\end{enumerate}

In the above relations, variables $x_i$ are given by the expressions
\begin{eqnarray}
 x_1 &=& {\cal{Y}}_{n,m} u_{n,m}+{\cal{X}}_{n,m} u_{n+1,m}\,, \quad x_2 \,=\, {\cal{Y}}_{n,m} u_{n+1,m}+{\cal{X}}_{n,m} u_{n,m}\,, \nonumber \\
 x_3 &=& {\cal{Y}}_{n,m} u_{n,m+1}+{\cal{X}}_{n,m} u_{n+1,m+1}\,, \quad x_4 \,=\, {\cal{Y}}_{n,m} u_{n+1,m+1}+{\cal{X}}_{n,m}\,, u_{n,m+1}\,,\label{3legvars}
\end{eqnarray}
or similar expressions by interchanging mutually $u_{n,m}$ with $u_{n+1,m+1}$, and $u_{n+1,m}$ with $u_{n,m+1}$. Relations (\ref{3legvars}) and our conventions imply that, $x_i$ are naturally adapted to the white diagonals, i.e $(x_1 ,x_4)$ corresponds to a white diagonal independently of the choice for $\sigma$.

On the other hand, $y_i$ follow from the corresponding $x_i$ by interchanging ${\cal{X}}_{n,m}$ with ${\cal{Y}}_{n,m}$. They are adapted to the black diagonals, since $(y_1, y_4)$ corresponds to a black diagonal independently of $\sigma$.

The above three-leg forms allow us to derive two different {\it{discrete Toda type equations}} from each one of the ${\rm H}_\epsilon^\sigma$ equations. The resulting Toda systems can be written in a compact form as\footnote{The resulting equation remains invariant under the interchange of $u_{n+2,m}$ and $u_{n,m+2}$.}
\be {\cal{H}}_{n,m}(u_{n,m},u_{n+2,m},u_{n+1,m+1})-{\cal{H}}_{n,m}(u_{n,m+2},u_{n+2,m+2},u_{n+1,m+1})=0\,,\label{defToda} \ee
where
\begin{enumerate}[$\bullet$]
\item for equations ${\rm H1}^\sigma_\epsilon$
\begin{equation}
 {\cal{H}}_{n,m}(x,y,z)\,=\,\frac{x-y}{(x\,-\,z) (y\,-\,z)\, -\, \epsilon\,{\cal{X}}_{n,m} (\alpha\,-\,\beta)^2}\,, \label{TodaH1}
\end{equation}
\item for equations ${\rm H2}^\sigma_\epsilon$
\begin{eqnarray}
 && {\cal{H}}_{n,m}(x,y,z)= \nonumber \\ 
 && \qquad \quad  \frac{x-y}{(x-z  + a_{n,m}) (y-z +b_{n,m})\, -\, 8\, \epsilon\,{\cal{X}}_{n,m}  (\alpha-\beta)^2 (z + \alpha (1-2 \epsilon \alpha))},  \label{TodaH2}
\end{eqnarray}
$\phantom{\bullet}$ with
$$a_{n,m} := (\beta-\alpha) (1 - 2\epsilon {\cal{X}}_{n,m} (\alpha+\beta))\,,$$ 
$$b_{n,m}:= (\alpha-\beta) (1 - 2 \epsilon {\cal{X}}_{n,m} ( 3 \alpha - \beta))\,,$$
\item and, for equations ${\rm H3}^\sigma_\epsilon$
\begin{equation}
 {\cal{H}}_{n,m}(x,y,z)\,=\,\frac{x-y}{\alpha \,\beta \,(\alpha \,x\,-\, \beta \,z)\, (\beta \, y\,-\, \alpha \,z)\, -\, \delta
\, \epsilon \,{\cal{X}}_{n,m} (\alpha^2\,-\,\beta^2)^2}\,,  \label{TodaH3}
\end{equation}
\end{enumerate}
respectively.

It is worth noting that, when ${\cal{X}}_{n,m} = 0$, all of the above Toda systems are identical to the corresponding Toda systems following from the original H equations of \cite{ABS}.

\section{B{\"a}cklund transformation and Lax pairs}

The interpretation of the multidimensional consistency property as a B{\"a}cklund transformation is well known, and leads to a Lax pair for the equation under consideration, \cite{BobSuris,Nij1}. In that sense, equations ${\rm H}^\sigma_\epsilon$ corresponding to the two different choices of $\sigma$ are related to each other by a B{\"a}cklund transformation, in terms of which Lax pairs are constructed for both of them. This section is devoted to the presentation of this transformation and its superposition principle, as well as of the Lax pairs.

\begin{thm}[B{\"a}cklund transformation] \label{propBac}
Let $u$ be a solution of the deformed equation
\be {\cal{E}}(u_{n,m},u_{n+1,m},u_{n,m+1},u_{n+1,m+1};n,m;\alpha,\beta;\sigma)\,=\,0\,. \label{autoBacdiseq1} \ee
Then, the function $v$ determined by the system
\be 
{\mathds{B}}(u,v,\lambda) \,:=\,\left\{ \begin{array}{l} {\cal{E}}(u_{n,m},u_{n+1,m},v_{n,m},v_{n+1,m};n,m;\alpha,\lambda;\sigma) \,=\, 0\\
{\cal{E}}(u_{n,m},u_{n,m+1},v_{n,m},v_{n,m+1};n,m;\beta,\lambda;\sigma) \,=\, 0 \end{array} \right.\,, \label{discautoBac}
\ee
is a solution of the equation
\be {\cal{E}}(v_{n,m},v_{n+1,m},v_{n,m+1},v_{n+1,m+1};n,m;\alpha,\beta;-\sigma)\,=\,0\,. \label{autoBacdiseq2} \ee

Conversely, if $v$ is a solution of (\ref{autoBacdiseq2}), then the function $u$ defined via the inverse of transformation (\ref{discautoBac}), i.e.
\be 
{\mathds{B}}(v,u,\lambda) \,:=\,\left\{ \begin{array}{l} {\cal{E}}(v_{n,m},v_{n+1,m},u_{n,m},u_{n+1,m};n,m;\alpha,\lambda;-\sigma) \,=\, 0\\
{\cal{E}}(v_{n,m},v_{n,m+1},u_{n,m},u_{n,m+1};n,m;\beta,\lambda;-\sigma) \,=\, 0 \end{array} \right.\,, \label{invdiscautoBac}
\ee
is a solution of equation (\ref{autoBacdiseq1}).

Thus, ${\mathds{B}}(u,v,\lambda)$ constitutes a B{\"{a}}cklund transformation between the deformed equations (\ref{autoBacdiseq1}), (\ref{autoBacdiseq2}).
\end{thm}

\begin{pr}
Follows from the proof of the corresponding proposition given in \cite{Xen,X1}.\hfill $\Box$
\end{pr}

\begin{figure}[h]
\begin{minipage}{14pc}
\centertexdraw{ \setunitscale .45 \linewd 0.01 \arrowheadtype t:F \arrowheadsize l:0.2 w:0.16
\move(0 0) \lvec (1.5 1.5) \lvec(3 0) \lvec(1.5 -1.5) \lvec(0 0) 
\move(0 0) \fcir f:0.0 r:0.08
\move(1.5 1.5) \fcir f:0.85 r:0.08
\move(3 0) \fcir f:0.0 r:0.08
\move(1.5 -1.5) \fcir f:0.85 r:0.08
\move(.8 .8) \avec(.85 .85)
\move(2.25 .75) \avec(2.3 .7)
\move(.8 -.8) \avec(.85 -.85)
\move(2.25 -.75) \avec(2.3 -.7) 
\htext (-0.5 -0.1) {$u^{0}$} \htext (1.4 1.65) {$v^1$} \htext (3.2 -.1) {$u^{12}$} \htext (1.4 -1.8) {$v^2$}
\htext(0.55 0.9){$\lambda_1$} \htext(0.55 -1.1){$\lambda_2$} \htext(2.2 0.9){$\lambda_2$} \htext(2.3 -1.1){$\lambda_1$}}
\caption{Bianchi diagram} \label{Bianchicomd}
\end{minipage}
\begin{minipage}{16pc}
\centertexdraw{ \setunitscale .3 \linewd 0.01 \arrowheadtype t:F \arrowheadsize l:0.2 w:0.16
\move(0 0) \lvec(1.5 1.5) \lvec(3 0) \lvec(1.5 -1.5) \lvec(0 0) \lfill f:0.65
\move(3 0) \lvec(4.5 1.5) \lvec(6 0) \lvec(4.5 -1.5) \lvec(3 0) \lfill f:0.65
\move(0 0) \lvec (3 3) \lvec(6 0) \lvec(3 -3) \lvec(0 0) 
\move(1.5 1.5) \lvec(4.5 -1.5) \move(1.5 -1.5) \lvec(4.42 1.42) \move(4.5 1.5) \fcir f:0.85 r:0.1
\move(0 0) \fcir f:0.0 r:0.1 \move(3 3) \fcir f:0.0 r:0.1 
\move(1.5 1.5) \fcir f:0.85 r:0.1 \move(6 0) \fcir f:0.0 r:0.1 
\move(3 0) \fcir f:0.0 r:0.1 \move(3 -3) \fcir f:0.0 r:0.1 
\move(1.5 -1.5) \fcir f:0.85 r:0.1 \move(4.5 -1.5) \fcir f:0.85 r:0.1 
\move(.8 .8) \avec(.85 .85) \move(2.3 2.3) \avec(2.35 2.35) \move(3.8 .8) \avec(3.85 .85)
\move(2.25 .75) \avec(2.3 .7) \move(3.75 2.25) \avec(3.8 2.2) \move(5.25 0.75) \avec(5.3 0.7)
\move(.8 -.8) \avec(.85 -.85) \move(2.3 -2.3) \avec(2.35 -2.35) \move(3.65 -2.35) \avec(3.7 -2.3)
\move(2.25 -.75) \avec(2.3 -.7) \move(3.7 -.7) \avec(3.75 -.75)  \move(5.15 -0.85) \avec(5.2 -0.8)
\htext(3 3.2){$u^1$} \htext(6.15 0){$\bar{u}$} \htext(4.7 1.5) {$v^{12}$} \htext(4.65 -1.65) {$v^{21}$} \htext(3 -3.35) {$u^{2}$}
\htext (-0.5 -0.1) {$u^{0}$} \htext (1.35 1.65) {$v^1$} \htext (2.7 -.7) {$u^{12}$} \htext (0.95 -1.8) {$v^2$}
\htext(0.55 0.95){$\lambda_1$} \htext(2.05 2.45){$\lambda_1$}  
\htext(0.55 -1.2){$\lambda_2$} \htext(2.05 -2.7){$\lambda_2$}  }
\caption{Double Bianchi diagram} \label{Bianchicomd5}
\end{minipage}
\end{figure}

\begin{thm}[Superposition principle (Bianchi commuting diagram)] \label{Bianchidiscrete}
Let $u^0$ be a solution of the deformed equation
\be {\cal{E}}(u_{n,m},u_{n+1,m},u_{n,m+1},u_{n+1,m+1};n,m;\alpha,\beta;\sigma)\,=\,0\,, \label{Bianchi1}\ee
and $v^1$, $v^2$ be the solutions of its counterpart
\be {\cal{E}}(v_{n,m},v_{n+1,m},v_{n,m+1},v_{n+1,m+1};n,m;\alpha,\beta;-\sigma)\,=\,0\,, \label{Bianchi2}\ee
generated by $u^0$ via the B{\"a}cklund transformation ${\mathds{B}}$ with parameters $\lambda_1$ and $\lambda_2$, respectively. Then, there is a new solution $u^{12}$ of equation (\ref{Bianchi1}), which is constructed according to the Bianchi commuting diagram, Figure \ref{Bianchicomd}, and is given algebraically by
\be {\cal{E}}\left(u^0,v^1,v^2,u^{12};n,m;\lambda_1,\lambda_2;\sigma \right)\,=\,0\,. \label{Bianchi4} \ee
\end{thm}

\begin{pr}
Follows from the proof of the corresponding proposition in \cite{Xen,X1}. \hfill $\Box$
\end{pr}

The algebraic formula (\ref{Bianchi4}) involves two solutions of the deformed equation (\ref{Bianchi1}) and two solutions of its counterpart (\ref{Bianchi2}). One may obtain algebraic formulas relating five solutions of the deformed equation (\ref{Bianchi1}) by applying twice the transformation ${\mathds{B}}$ and using Proposition \ref{Bianchidiscrete} in accordance with Figure \ref{Bianchicomd5}. This procedure can be straightforwardly materialized using the corresponding three-leg forms given in the previous section leading to
$${\cal{H}}_{n,m}(u^0,u^i,u^{12})-{\cal{H}}_{n,m}(u^j,\bar{u},u^{12})=0\,,\quad i \ne j\,,\quad i,\,j=1,2\,,$$
where the corresponding functions ${\cal{H}}_{n,m}$ are given in (\ref{TodaH1})-(\ref{TodaH3}) with $(\alpha,\beta) \longrightarrow (\lambda_1,\lambda_2)$.

Using the fact that, a B{\"a}cklund transformation may be regarded as the gauge transformation converting the matrices of the Lax pair into upper triangular ones, \cite{Crampin}, one may derive a Lax pair for the equations under consideration, \cite{Xen,X1}. The result is given in the next proposition, where we use the notation ${\cal{E}}_{,i}$ to denote the derivative of ${\cal{E}}$ with respect to its $i$-th argument.
\begin{thm}[Lax pair]
The deformed equation
$${\cal{E}}(u_{n,m},u_{n+1,m},u_{n,m+1},u_{n+1,m+1};n,m;\alpha,\beta;\sigma)\,=\,0$$
is the compatibility condition of the linear system
\begin{subequations}\label{Laxpairall}
\be
\begin{array}{l} \Psi_{n+1,m} = {\cal{L}}_{n,m}\left(u_{n,m},u_{n+1,m};\alpha,\lambda;\sigma\right) \Psi_{n,m} \\
\Psi_{n,m+1} = {\cal{L}}_{n,m}\left(u_{n,m},u_{n,m+1};\beta,\lambda;\sigma \right) \Psi_{n,m}\end{array}\,,
\ee
where
\be
 {\cal{L}}_{n,m}(x_1,x_2;\alpha,\lambda;\sigma)\,:=\,\frac{1}{\sqrt{k(\alpha,\lambda)\,f(x_1,x_2,\alpha)}} \left(\begin{array}{c c} {\cal{E}}_{,4} & - {\cal{E}}_{,3 4} \\ {\cal{E}} & -{\cal{E}}_{,3} \end{array} \right)\label{Laxpairall2}
\ee
and ${\cal{E}}\,=\,{\cal{E}}(x_1,x_2,x_3,x_4;n,m;\alpha,\lambda;\sigma)$ and its derivatives are evaluated at $x_3 = x_4 = 0$.
\end{subequations}
\end{thm}

The matrix on the right hand side of definition (\ref{Laxpairall2}) has the following explicit form for each one of the deformed equations. 
\begin{enumerate}[a)]
\item Equations ${\rm H1}_\epsilon^\sigma$
\be 
\left(\begin{array}{cc} -x_2 & -1 \\ x_1 x_2 - \alpha + \lambda & x_1  \end{array} \right)\, +\, \epsilon\, (\alpha-\lambda)\, \left( \begin{array}{cc} {\cal{Y}}_{n,m} x_1 & 0 \\ 0 & -{\cal{X}}_{n,m} x_2 \end{array} \right)
\ee
\item Equations ${\rm H2}_\epsilon^\sigma$
$$
\left(\begin{array}{cc} -x_2 -\alpha + \lambda & -1 \\ (x_1+\lambda) (x_2+\lambda) - \alpha (x_1+x_2) - \alpha^2 & x_1 +\alpha-\lambda  \end{array} \right) \hspace{5cm} $$
\be    +\,2\, \epsilon \,(\alpha-\lambda) \left( \begin{array}{cc} {\cal{Y}}_{n,m} (x_1+\alpha+\lambda) & 0 \\ (\alpha+\lambda) ({\cal{Y}}_{n,m} x_1 + {\cal{X}}_{n,m} x_2) + \alpha^2 + \lambda^2 & -{\cal{X}}_{n,m} (x_2+\alpha+\lambda) \end{array} \right)
\ee
\item Equations ${\rm H3}_\epsilon^\sigma$
\begin{equation}
 \left( \begin{array}{cc} -\lambda x_2 & -\alpha \\ \alpha x_1 x_2 + \delta^2 (\alpha^2-\lambda^2) & \lambda x_1 \end{array} \right) -\, \frac{\epsilon (\alpha^2-\lambda^2)}{\alpha \lambda} \left(\begin{array}{cc} {\cal{Y}}_{n,m} x_1 & 0 \\ 0 & - {\cal{X}}_{n,m} x_2 \end{array} \right)
\end{equation}
\end{enumerate}

\begin{rem}
The explicit forms of two successive Lax matrices, e.g. 
$$ {\cal{L}}_{n,m}(u_{n,m},u_{n+1,m};\alpha,\lambda;\sigma)\quad {\mbox{and}} \quad {\cal{L}}_{n,m+1}(u_{n,m+1},u_{n+1,m+1},\alpha,\lambda;\sigma)\,,$$
are related to each other as
$${\cal{L}}_{n,m+1}(u_{n,m+1},u_{n+1,m+1},\alpha,\lambda;\sigma)\,=\,-\,\Big\{ {\cal{L}}_{n,m}(u_{n+1,m+1},u_{n,m+1};\alpha,\lambda;-\sigma) \Big\}^{-1}\,.$$
\end{rem}

\section{Symmetry analysis}

This section contains the results on the symmetry analysis of the deformed equations, the derivation of which follows from the corresponding ones presented in \cite{TTX} and \cite{X1}. In particular, every deformed equation admits a pair of three point generalized symmetries, as well as a pair of extended generalized, the generators of which are determined completely by the polynomial $f$. Moreover, infinite hierarchies of generalized symmetries are constructed, the members of which are determined inductively.

\begin{thm}
Every deformed equation 
$${\cal{E}}(u_{n,m},u_{n+1,m},u_{n,m+1},u_{n+1,m+1};n,m;\alpha,\beta;\sigma)\,=\,0 $$
admits two local three-point generalized symmetries with generators the vector fields
\begin{subequations} \label{genformgensym}
\be  {\mathbf{v}}_n = {\rm R}_n^{[0]} \partial_{u_{n,m}} := \left( \frac{f(u_{n,m},u_{n+1,m};\alpha)}{u_{n+1,m}-u_{n-1,m}} - \frac{1}{2} f_{,u_{n+1,m}}(u_{n,m},u_{n+1,m};\alpha) 
 \right) \partial_{u_{n,m}}, \ee
and
\be  {\mathbf{v}}_m = {\rm R}_m^{[0]}\,\partial_{u_{n,m}}\,:=\, \left( \frac{f(u_{n,m},u_{n,m+1};\beta)}{u_{n,m+1}-u_{n,m-1}} - \frac{1}{2} 
f_{,u_{n,m+1}}(u_{n,m},u_{n,m+1};\beta) \right) \partial_{u_{n,m}}\,,\ee
\end{subequations}
respectively.

Moreover, the generator of any five point extended generalized symmetry of a deformed equation is necessarily a vector field of the form
\begin{eqnarray*}
{\bf v} &=& a(n;\alpha,\beta) {\mathbf{v}}_n + b(m;\alpha,\beta) {\mathbf{v}}_m + \frac{1}{2}\,\psi(n,m,u_{n,m};\alpha,\beta) \partial_{u_{n,m}}\\
 & &+ \,\xi(n,m;\alpha,\beta) \partial_\alpha \,+\, \zeta(n,m;\alpha,\beta) \partial_\beta \,,
 \end{eqnarray*}
where the functions $a(n;\alpha,\beta)$, $b(m;\alpha,\beta)$, $\psi(n,m,u_{n,m};\alpha,\beta)$, $\xi(n,m;\alpha,\beta)$ and $\zeta(n,m;\alpha,\beta)$ are determined through equation (\ref{eq:infcr}).
\end{thm}

Applying the above analysis to each one of the deformed equations, one is led to the following results. 
\begin{enumerate}
\item The only deformed equations admitting point symmetries are
\begin{enumerate}
\item Equations ${\rm H1}^\sigma_\epsilon$, admitting an extended point symmetry with generator the vector field ${\bf x} = \partial_\alpha+\partial_\beta$, and
\item Equations ${\rm H3}^\sigma_\epsilon$ for $\delta=0$, which admit a local point symmetry with generator the vector field $ {\bf x} =u_{n,m} \partial_{u_{n,m}}$.
\end{enumerate}
\item All the deformed equations admit the local three-point generalized symmetries generated by the vector fields ${\bf v}_n$, ${\bf v}_m$ and no other symmetry of this kind.
\item Only equations ${\rm H1}^\sigma_\epsilon$ admit additionally a local five-point generalized symmetry generated by the vector field $n {\bf v}_n + m {\bf v}_m$.
\item All the deformed equations admit a pair of extended three-point generalized symmetries with generators the vector fields
$${\bf V}_n\,=\, n\,{\bf v}_n\,-\,r(\alpha)\,\partial_\alpha\,,\quad {\bf V}_m\,=\,m\,{\bf v}_m\,- \,r(\beta)\,\partial_\beta\,, $$
respectively, where $r(x)=1$ for ${\rm H1}_\epsilon^\sigma$ and ${\rm H2}_\epsilon^\sigma$, and $r(x)=-x/2$ for ${\rm H3}_\epsilon^\sigma$.
\end{enumerate}

As in the case of the ABS equations, \cite{X1}, the extended generalized symmetries ${\bf V}_n$, ${\bf V}_m$ are master symmetries for the symmetry generators ${\bf v}_n$, ${\bf v}_m$. That fact leads to the infinite hierarchies of generalized symmetries
\begin{subequations}\label{infhierrec}
\be  {\bf v}_i^{[k+1]}\,=\,{\rm R}_i^{[k+1]}\,\partial_{u_{n,m}}\,:=\,\left[ {\bf V}_i,{\bf v}_i^{[k]}\right]\,,\quad k\,=\,0,\,1,\,\ldots,\quad{\mbox{and}}\quad i\,=\,n,\,m\,,\ee
the first members of which are ${\bf{v}}_n^{[0]} = {\bf{v}}_n$ and ${\bf{v}}_m^{[0]} = {\bf{v}}_m$, respectively.

The characteristic ${\rm R}_i^{[k]}$ involves the values of $u$ at $(2 k + 3)$ points in the $i$ direction of the lattice and is determined by applying successively the linear differential operator 
\be  {\cal{R}} = \sum_{\ell=-\infty}^\infty \ell \, \left({\cal{S}}_n^{(\ell)} {\rm R}_n^{[0]}\right) \partial_{u_{n+\ell,m}} + \sum_{\ell=-\infty}^\infty \ell\,\left({\cal{S}}_m^{(\ell)} {\rm R}_m^{[0]}\right) \partial_{u_{n,m+\ell}} -r(\alpha)\,\partial_\alpha - r(\beta)\,\partial_\beta
 \label{recoper} \ee
on ${\rm R}_i^{[0]}$, i.e.
\be {\rm R}_i^{[k]}\,=\,{\cal{R}}^{\,k} {\rm R}_i^{[0]}\,,\quad k\,=\,0,\,1\,\ldots,\quad{\mbox{and}}\quad i\,=\,n,\,m\,. \label{infhierrec3} \ee
\end{subequations}

\begin{rem}
The operator ${\cal{R}}$ may be regarded as recursion operator for equations ${\rm H2}_\epsilon^\sigma$ and ${\rm H3}_\epsilon^\sigma$, since it maps the local generalized symmetries ${\bf v}_n$, ${\bf v}_m$ to higher local generalized symmetries.\hfill $\Box$
\end{rem}

\section{Conclusions and perspectives}
We have presented the integrability aspects and the symmetry analysis of the deformations of the H list. B{\"a}cklund transformations and Lax pairs for all of the deformed equations were derived from their multidimensional consistency. Black and white centered three-leg forms were constructed, leading to deformations of discrete Toda type systems. Their symmetry analysis was presented, leading to infinite hierarchies of generalized symmetries.

It would be interesting to study solutions of the deformed equations. In this direction, one could employ their symmetries in order to derive similarity solutions, and continuous symmetric solutions in the spirit of \cite{TX}. Regarding to the latter, such reductions will also provide a link of the deformed equations to integrable systems of partial differential equations. Additionally, it would be interesting to study finite dimensional mappings resulting from periodic reductions of the deformed equations.

\section*{Acknowledgments}
The paper was completed while the authors were both visiting the Isaac Newton Institute for Mathematical Sciences (INI), Cambridge, UK, as part of the programme {\it{Discrete Integrable Systems}} (January-June 2009). They wish to thank the programme organisers for the invitations and the INI for hospitality.

\appendix

\section{The polynomials of the deformed equations}
Polynomials $G_2$ follow from polynomials $G_1$ by substituting ${\cal{Y}}_{n,m}$ with ${\cal{X}}_{n,m}$.
\begin{enumerate}[$\bullet$]
\item Equations ${\rm H1}^\sigma_\epsilon$
\begin{eqnarray*}  
&& k(\alpha,\beta) = \beta-\alpha\,,\quad f(u,x;\alpha) =  1 - \epsilon \left( {\cal{Y}}_{n,m} \,u^2 + {\cal{X}}_{n,m}\, x^2 \right)\\
&& G_1(x,y) = (x-y)^2 - \epsilon (\alpha-\beta)^2 {\cal{Y}}_{n,m} 
\end{eqnarray*}
\item Equations ${\rm H2}^\sigma_\epsilon$
\begin{eqnarray*}
&& k(\alpha,\beta)= \beta-\alpha\,,\quad f(u,x;\alpha) = 2 \big\{ u + x + \alpha - 2 \epsilon \left[{\cal{Y}}_{n,m} (u+\alpha)^2 + {\cal{X}}_{n,m} (x+\alpha)^2\right]\big\}\\
&& G_1(x,y) = (x - y)^2 - (\alpha - \beta)^2 \big\{4\, \epsilon \,{\cal{Y}}_{n,m} \,(x+y-\epsilon (\alpha-\beta)^2) + 1 \big\} 
\end{eqnarray*}
\item Equations ${\rm H3}^\sigma_\epsilon$
\begin{eqnarray*}
&& k(\alpha,\beta) = \alpha^2 - \beta^2\,,\quad f(u,x;\alpha) = u x + \delta \alpha \,-\, \frac{\epsilon}{\alpha} \left( {\cal{Y}}_{n,m} \,u^2\,+\,{\cal{X}}_{n,m}\, x^2\right) \\
&& G_1(x,y) = (\alpha x - \beta y) (\beta x - \alpha y)  - {\cal{Y}}_{n,m}\,\frac{\epsilon\,\delta (\alpha^2-\beta^2)^2}{\alpha \beta}
\end{eqnarray*}
\end{enumerate}


\end{document}